\documentclass{ieeeaccess}
\usepackage{cite}
\usepackage{amsmath,amssymb,amsfonts}
\usepackage{algorithmic}
\usepackage{graphicx}
\usepackage{textcomp}

\usepackage{booktabs}
\usepackage{mathtools}
\usepackage{xurl,lineno,microtype}

\usepackage{bm}
\makeatletter
\AtBeginDocument{\DeclareMathVersion{bold}
\SetSymbolFont{operators}{bold}{T1}{times}{b}{n}
\SetSymbolFont{NewLetters}{bold}{T1}{times}{b}{it}
\SetMathAlphabet{\mathrm}{bold}{T1}{times}{b}{n}
\SetMathAlphabet{\mathit}{bold}{T1}{times}{b}{it}
\SetMathAlphabet{\mathbf}{bold}{T1}{times}{b}{n}
\SetMathAlphabet{\mathtt}{bold}{OT1}{pcr}{b}{n}
\SetSymbolFont{symbols}{bold}{OMS}{cmsy}{b}{n}
\renewcommand\boldmath{\@nomath\boldmath\mathversion{bold}}}
\makeatother

\def\BibTeX{{\rm B\kern-.05em{\sc i\kern-.025em b}\kern-.08em
    T\kern-.1667em\lower.7ex\hbox{E}\kern-.125emX}}

\DeclarePairedDelimiter\bra{\langle}{\rvert}
\DeclarePairedDelimiter\ket{\lvert}{\rangle}
\DeclarePairedDelimiterX\braket[2]{\langle}{\rangle}{#1\,\delimsize\vert\,\mathopen{}#2}

\DeclareMathOperator*{\argmin}{arg\,min}

\newcommand{\minus}{\scalebox{0.75}[1.0]{$-$}}

\begin{document}
\history{}
\doi{}

\title{Quantum Optimization Methods for Satellite Mission Planning}
\author{\uppercase{Ant\'on Makarov}\authorrefmark{1,2}, \uppercase{Carlos P\'erez-Herrad\'on}\authorrefmark{1}, \uppercase{Giacomo Franceschetto}\authorrefmark{3}, \uppercase{M\'arcio M. Taddei}\authorrefmark{3}, \uppercase{Eneko Osaba}\authorrefmark{4}, \uppercase{Paloma del Barrio}\authorrefmark{1}, \uppercase{Esther Villar-Rodriguez}\authorrefmark{4}, \uppercase{Izaskun Oregi}\authorrefmark{4,5}}

\address[1]{GMV, Calle Isaac Newton 11, 28760 Tres Cantos, Madrid, Spain}
\address[2]{Institute of Fundamental Physics, IFF-CSIC, Calle Serrano 113B, 28006 Madrid, Spain}
\address[3]{ICFO - Institut de Ciencies Fotoniques, The Barcelona Institute of Science and Technology, 08860 Castelldefels, Barcelona, Spain}
\address[4]{TECNALIA, Basque Research and Technology Alliance (BRTA), 48160 Derio, Spain}
\address[5]{European University of Gasteiz, EUNEIZ, 01013 Vitoria-Gasteiz, Spain}

\tfootnote{This work was supported by the Spanish Ministry of Science and Innovation under the Recovery, Transformation and Resilience Plan (CUCO, MIG-20211005), the Government of Spain (Severo Ochoa CEX2019-000910-S, FUNQIP, and European Union NextGenerationEU PRTR-C17.I1), the European Union (PASQuanS2.1, 101113690), GMV, Fundaci\'o Cellex, Fundaci\'o Mir-Puig, Generalitat de Catalunya (CERCA program). This work was also supported by the Basque Government through Plan complementario comunicaci\'on cu\'antica (EXP. 2022/01341) (A/20220551). G.F. acknowledges support from a ”la Caixa” Foundation (ID 100010434) fellowship. The fellowship code is LCF/BQ/DI23/11990070.}

\markboth
{Makarov \headeretal: Quantum Optimization Methods for Satellite Mission Planning}
{Makarov \headeretal: Quantum Optimization Methods for Satellite Mission Planning}

\corresp{Corresponding author: Carlos P\'erez-Herrad\'on (email: cperez.h@gmv.com).}

\begin{abstract}
Satellite mission planning for Earth observation satellites is a combinatorial optimization problem that consists of selecting the optimal subset of imaging requests, subject to constraints, to be fulfilled during an orbit pass of a satellite. The ever-growing amount of satellites in orbit underscores the need to operate them efficiently, which requires solving many instances of the problem in short periods of time. However, current classical algorithms often fail to find the global optimum or take too long to execute. Here, we approach the problem from a quantum computing point of view, which offers a promising alternative that could lead to significant improvements in solution quality or execution speed in the future. To this end, we study a planning problem with a variety of intricate constraints and discuss methods to encode them for quantum computers. Additionally, we experimentally assess the performance of quantum annealing and the quantum approximate optimization algorithm on a realistic and diverse dataset. Our results identify key aspects like graph connectivity and constraint structure that influence the performance of the methods. We explore the limits of today's quantum algorithms and hardware, providing bounds on the problems that can be currently solved successfully and showing how the solution degrades as the complexity grows. This work aims to serve as a baseline for further research in the field and establish realistic expectations on current quantum optimization capabilities.
\end{abstract}


\titlepgskip=-21pt

\maketitle

\section{Introduction}
\label{sec:introduction}

\PARstart{T}{he} Satellite Mission Planning Problem (SMPP) is a critical issue in the aerospace sector \cite{zhang2021mission}. Satellite operators face the task of determining the optimal subset of images to be captured during a satellite's orbital pass, based on a set of client requests and a defined value metric. The challenge is compounded by constraints such as geographical proximity incompatibilities, onboard storage limitations, and specific image configuration requirements, which make it unfeasible to collect all images. Fig. \ref{fig:smpp} depicts a simple scenario of a SMPP where the satellite has to choose from the set of all requested images (rectangles) which ones to take (green) and which to leave unattended (red). The problem’s complexity increases when considering the continuous influx of requests, thereby necessitating frequent plan updates and repeated execution of the planning algorithm within short time frames. This requirement underscores the importance of execution speed. Current industry challenges such as extending the problem to satellite constellations, incorporating additional constraints like cloud coverage, or real-time onboard re-planning \cite{zhang2021mission}, increase computational demands even further.

\Figure[!t]()[width=0.95\columnwidth]{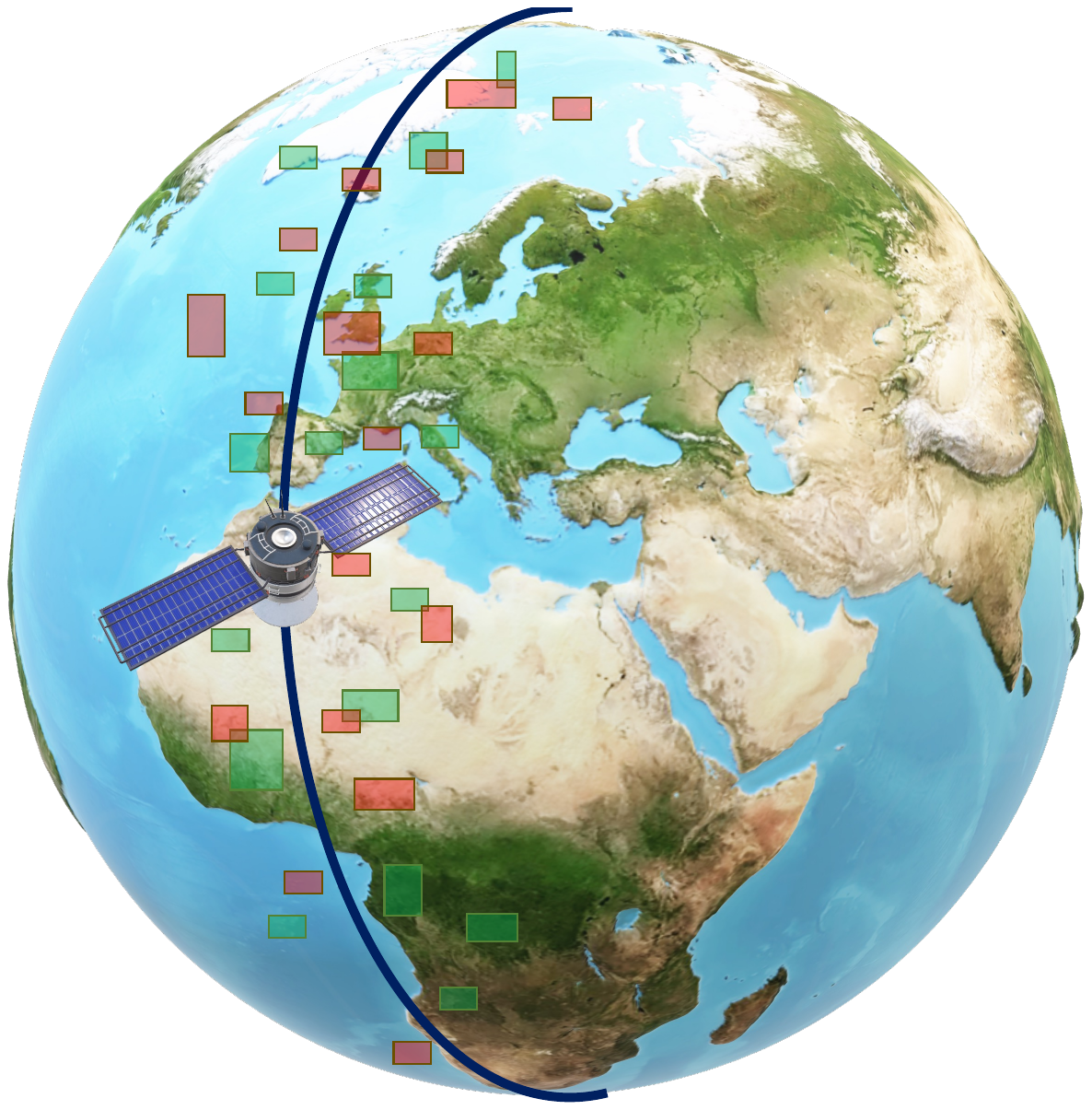}{SMPP diagram. A satellite is orbiting the Earth and has to choose which image requests, represented by rectangles, to capture. One possible solution to the planning problem is to take the green ones and discard or delay for the next orbital pass the red ones. 
\label{fig:smpp}}

The SMPP and its extensions exhibit high combinatorial complexity, as even its simpler versions can be cast to variants of the knapsack problem \cite{bensana1999earth}, which is known to be NP-complete \cite{karp1975computational}. This poses significant challenges for classical computing methods, even for moderately sized instances with few constraints. Traditional classical approaches have employed exact integer linear programming algorithms based on branch-and-bound/cut methods \cite{bensana1999earth, vasquez2003upper, ribeiro2010strong}, (meta-)heuristic algorithms \cite{Wei1399860tabu, barkaoui2020new, bianchessi2007heuristic, jiang2022dual}, and machine learning based techniques \cite{WANG20223478}. The latter are typically favored due to their ability to handle large problems and comply with execution-time constraints.

Quantum computing (QC) is emerging as a promising alternative for addressing hard optimization problems by leveraging properties of quantum physics \cite{symons2023practitioner, li2020quantum}. Several approaches are currently being investigated by the community, of which quantum annealing (QA) \cite{kadowaki1998quantum} and variational quantum algorithms such as the quantum approximate optimization algorithm (QAOA) \cite{farhi2014quantum} are among the most prominent. These methods, coupled with the appropriate hardware, may provide near- to mid-term benefits such as substantial computational speedups, better-quality solutions or a reduction in energy consumption.

In this paper, we study the SMPP from the QC point of view, using approaches from two of the leading and available paradigms: quantum annealers and gate-based devices. To do that, we present a formulation in which a variety of hard constraints are analyzed, paying extra attention to the more challenging ones. We use 31 different problem instances to evaluate our formulation as well as the performance of quantum annealing and QAOA methods on currently available quantum hardware and simulators. Six of these instances were directly sourced from the well-known SPOT5 dataset \cite{bensana1999earth}, while the remaining 25 were generated based on SPOT5 using an instance reductor specifically implemented for this study.

The rest of the paper is structured as follows. Section \ref{sec:background} provides a comprehensive background related to the problem under investigation. The mathematical models developed to efficiently formulate the problem from the classical and quantum points of view are presented in Section \ref{sec:modelling}. Section \ref{sec:methods} describes the quantum algorithms evaluated and the criteria for performance assessment and comparison. Section \ref{sec:expres} details the dataset, experimental design, conducted tests, and a discussion of the findings. The paper culminates with Section \ref{sec:conclusion}, which draws conclusions and outlines potential avenues for future research.

\section{Background}
\label{sec:background}

For years, quantum technologies have had a significant influence in the space sector \cite{kaltenbaek2021quantum}, with many resources dedicated to secure satellite communications and QKD \cite{vallone2015experimental, liao2017satellite} or quantum sensing \cite{MULLER2020105110}. Notably, there has been a recent broadening of focus within the industry towards quantum machine learning and optimization \cite{combarro2023practical}. This has led several major space agencies to establish initiatives like the ESA's Quantum Computing for Earth Observation\footnote{\url{https://eo4society.esa.int/projects/qc4eo-study/}} (QC4EO), NASA's Quantum Artificial Intelligence Laboratory\footnote{\url{https://www.nasa.gov/intelligent-systems-division/discovery-and-systems-health/nasa-quail/}} (QuAIL) or DLR's Quantum Computing Initiative\footnote{\url{https://qci.dlr.de/en/start/}} (QCI) to propel advancements in research and development within this domain.

Due to the importance of the SMPP in the aerospace industry and the limitations in terms of computational speed and solution quality when classical computers are used to solve it, methods from quantum computing appear as an attractive alternative. However, most of the published work does not deal with the SMPP per se, but with other related topics such as satellite routing, debris removal, or constellation optimization \cite{ORBi-eceda854-1443-4d98-8492-a10520f1a2d3, bass2018heterogeneous}. We take the scarcity of literature as motivation for this research and devote the rest of this section to analyzing the published related work.

The seminal work in this field was published in 2020 by Stollenwerk \emph{et al.} \cite{stollenwerk2020image,stollenwerk2021agile}, which deals for the first time with the SMPP using the quantum annealing paradigm. They investigate the potential and maturity of then-current quantum computers to solve real-world problems by carrying out an experimental study on a reduced number of small instances of the SMPP.

Another notable contribution is \cite{rainjonneau2023quantum}, where the authors demonstrate potential for quantum advantage by using a hybridized quantum-enhanced reinforcement learning algorithm and compare it to greedy and classical optimization algorithms on a satellite mission planning problem. On another note, in \cite{quetschlich2023hybrid} variational quantum algorithms such as the variational quantum eigensolver \cite{peruzzo2014variational} and QAOA are used to solve several simple instances of the SMPP with and without simulated noise. 

Additionally, the authors of the present paper published a preliminary study in \cite{makarov2023optimization}, which analyzed the problem formulation efficiency together with the performance of the state-of-the-art quantum annealing solvers at the time.

Thus, this work builds on that preliminary research and improves it by studying a larger set of problems with more diverse constraints. The study includes a theoretical complexity analysis of the models and a more in-depth experimental study using gate-based and annealing methods. More concretely, the main elements of this paper with respect to the state of the art and our preliminary findings can be summarized as follows:
\begin{itemize}
    \item The dataset employed in our research, taken from realistic simulations of a satellite mission, has been extended to cover a wide range of problem sizes and constraint types for a more in-depth study.
    \item Several types of challenging constraints are addressed in the study. Notably n-ary (capacity) and ternary (data flow) constraints, which greatly affect the performance of the algorithms, require ancillary qubits and a careful selection of the encoding method to model them for quantum computers.
    \item We analytically compare the complexity of encoding the constraints via several methods and propose a final, efficient formulation for the problem.
    \item We perform an experimental study on the performance of several approaches, namely, two purely quantum QA solvers, a quantum-classical hybrid method based on QA, and a gate-based QAOA that involves a classical optimization stage.
    \item We perform a detailed result analysis, measuring the quality of the solutions via the approximation ratio against reference classical solutions on the actual optimization function value, rather than using proxies such as the energy or only the best solution.
\end{itemize}
Lastly, it is noteworthy to acknowledge the existence of publications such as \cite{zhang2018solving, zhi2021variable}, which employ quantum-inspired genetic algorithms. These methodologies belong to the field of quantum-inspired evolutionary computation, which are classical algorithms augmented by principles derived from quantum physics for their design and thus are outside the scope of this paper.

\section{Modelling}
\label{sec:modelling}

In this section, we present the problem and the mathematical formulations used in this paper. First, in Section \ref{sec:problem} we introduce the key aspects of the problem we are going to solve. Then, in Section \ref{sec:classical}, we present a classical mathematical model for it. Subsequently, for each type of constraint we encounter in our model, we discuss in Section \ref{sec:qubo} possible penalty encodings and indicate their efficiency in terms of additional variables and quadratic terms. Finally, using these derivations we arrive at our formulation for the experiments which will be conducted in Section \ref{sec:expres}.

\subsection{Problem Description}
\label{sec:problem}

In this paper, we work with data from the SPOT5 satellite, which has three cameras on board and can satisfy two types of requests: mono images, for which only one of the cameras must be used, and stereo images, for which two of the cameras are required to produce the image. Each request has a weight that represents its value, and it may also have a capacity weight if it needs to be stored on the satellite's disk. The capacity value can vary depending on the camera used to take the request. This definition is subject to the following constraints:
\begin{enumerate}\renewcommand{\theenumi}{\alph{enumi}}
    \item Constraints enforcing that each request can only be collected once.
    \item Binary constraints that represent incompatibilities due to geographical proximity among certain pairs $(p, q)$, where $p$ and $q$ are pairs of (request, camera). We define, for each instance of the dataset, a set $S_2$ containing these forbidden pairs.
    \item Ternary constraints that represent the instantaneous data flow restrictions of the instruments, prohibiting taking more than 2 out of 3 requests at once for certain triples $(p,q,r)$, where $p,q$ and $r$ are pairs of (request, camera). We define, for each instance of the dataset, a set $S_3$ containing these forbidden triples.
    \item N-ary (capacity) constraints that represent the fact that the satellite has a limited amount of disk space available to store images before relaying them back to earth. This type of constraint is present only in some of the instances.
\end{enumerate}

For a more in-depth description of the dataset, the reader can refer to \cite{bensana1999earth}.

\subsection{Classical Model}
\label{sec:classical}

The classical formulation of the SMPP used in this paper builds upon the ideas provided on our previous research \cite{makarov2023optimization}, and it is stated using mathematical programming as follows:

Let $x_{i,j}$ be the binary decision variables, defined as:
\begin{equation}
x_{i,j}=
    \begin{cases}
        1 & \text{if request $i$ is taken with camera $j$},\\
        0 & \text{otherwise},
    \end{cases}
\end{equation}
where $i \in \{1, 2, \ldots, N\}$ is the index representing the image requests, $N$ being the total amount of requests and $j \in \{1, 2, 3, 4\}$ the identifier of the camera. There are three physical cameras that can take mono images and we define camera 4 to represent the combined use of cameras 1 and 3 to take stereo images. The necessary variables are created based on the specific data instance, following the notation mentioned above. The objective function to be optimized can then be defined as:
\begin{equation}
F(x) = \sum_i \sum_j w_i x_{i,j},
\end{equation}
where $w_i \in \mathbb{R}_{\geq 0}$ is the weight or value of fulfilling request $i$. Note that although our task is to maximize the value, we can express it as the minimization of the negative of the objective function. This optimization is subject to the following constraints:
\begin{subequations}
\begin{align}
\sum_j x_{i,j} &\leq 1, \quad \forall i, \label{const:once} \\
x_{p} + x_{q} &\leq 1, \quad \forall \left(p,q\right) \in S_2, \label{const:pairs}\\
x_{p} + x_{q} + x_{r} &\leq 2, \quad \forall \left(p, q, r\right) \in S_3, \label{const:ternary} \\
\sum_i\sum_j c_{i,j} x_{i,j} &\leq C, \label{const:capacity}\\
x_{i,j} &\in \{0, 1\}, \label{const:binary}
\end{align}\label{eq:constraints}
\end{subequations}\\
where constraints \eqref{const:once}-\eqref{const:capacity} refer to the four constraints mentioned above, respectively. Here $C \in \mathbb{Z}_{\geq 0}$ is the total capacity of the disk and $c_{i,j} \in \mathbb{R}_{\geq 0}$ the amount of space a request $i$ uses when taken with camera $j$. For some $(i,j)$ we effectively have $c_{i,j}=0$ due to the corresponding orbit points being close enough to a ground station for immediate relay, hence no storage requirement. Importantly, this makes constraint \eqref{const:capacity} take quite different forms for different instances. As a final remark, note that we can always flatten our decision variables into a one dimensional vector through the map $x_{i,j} \mapsto x_{i'}$, which is what we will use in practice for our implementation of the experiments.

\subsection{Efficient Quantum Model}
\label{sec:qubo}

Quantum optimization methods usually require to express the problem in a quadratic unconstrained binary optimization (QUBO) formulation due to their connection to the Ising model \cite{lucas2014ising} and adiabatic quantum computing \cite{albash2018adiabatic}, which serves as inspiration for many current methods like QA and QAOA. However, mapping a classical optimization problem to a QUBO is, in general, a difficult task that the quantum optimization community is commonly faced with \cite{glover2018tutorial}. Additionally, the efficiency of this mapping is critical to the performance of current quantum computers, given  their limitations in scale, connectivity, and stability. To this end, we will focus our attention on the encoding of constraints as penalties. In a nutshell, penalties are a method of encoding constraints in the objective function by adding a large $M\in \mathbb{R}$ multiplied by an at most quadratic polynomial that represents the constraint. Thus, on a minimization problem, adding a large positive $M$ makes the objective function larger when the constraint is violated. We analyze the efficiency of the encodings by looking at the two factors that are within our control and impact the problem size the most: the number of slack variables (ancillary qubits) and the number of quadratic interaction terms (qubit couplings) introduced by the encoding methods.

Constraints \eqref{const:once} can be trivially converted to penalties by taking the sum of the two-by-two products of the variables involved in each constraint, resulting in no variable overhead and $\binom{k}{2}=\mathcal{O}(k^2)$ quadratic interactions, where $k$ is the number of summation terms. In our case, $k \leq 3$ since the image is either stereo (only possible to take the image with camera 4) or mono (possible to take the image with at most three different cameras). Therefore, at most three interaction terms are introduced for each constraint. We proceed in the same fashion with constraints \eqref{const:pairs}, as they have a similar structure. In this case, $k=2$ and thus no additional terms are introduced and one interaction term is added for each constraint.

Constraint \eqref{const:ternary} can be efficiently treated in a custom manner. First, we write an equivalent cubic penalty term $M \ x_{p} x_{q} x_{r}$ and then reduce it to quadratic following Boros \emph{et al.} \cite[Sec 4.4]{Boros2002}. A single slack variable $s_1$ replaces the pair $x_{q} x_{r}$, and one extra interaction term is added, resulting in the following quadratic penalty,
\begin{equation}
M \ x_p s_1 + M ( x_qx_r-2x_qs_1-2x_rs_1+3s_1).
\label{eq:penalty3cam}
\end{equation}

This method has advantages for constraints such as \eqref{const:ternary} because it involves fewer terms than the more general binary expansion (discussed next) and introduces a single slack variable per constraint. Furthermore, if the pair $x_q x_r$ is present in multiple constraints, it can be replaced by the same slack variable in all of them, preventing the introduction of additional slack variables.

To deal with constraint \eqref{const:capacity}, which, as mentioned above, can take different forms depending on the data, we resort to the standard binary expansion. For ease of reading, we take $p=(i,j)$ and rewrite the constraint as:
\begin{equation}
    \sum_{p} c_px_p \leq C,
    \label{eq:generic_constraint}
\end{equation}
where we sum over every pair $p$ for which $c_p\neq0$, with $P$ such pairs in total. Now we transform this inequality into an equality. To this end, we add binary slack variables $s_d$ such that the binary representation reaches or exceeds $C$, that is, we find the minimum $D$ such that $\sum_{d=1}^D 2^{d-1} \geq C$. This results in an overhead of $D = \lceil \log_2 C \rceil$ slack variables. The constraint can now be written as:
\begin{equation}
    \sum_p c_px_p + \sum_{d=1}^{D} 2^{d-1}s_d = C,
    \label{eq:generic_constraint_slacks}
\end{equation}
which allows the use of the general penalty method:
\begin{equation}
    M\left(\sum_p c_px_p + \sum_{d=1}^{D} 2^{d-1}s_d - C\right)^2.
    \label{eq:generic_constraint_expansion}
\end{equation}
Therefore, with this method we are adding $\mathcal{O}(\log C)$ ancilla qubits and $\mathcal{O}\left((P+\log C)^2\right)$ quadratic interactions for each constraint of that type.

As a note, although constraint \eqref{const:ternary} is a special case of \eqref{eq:generic_constraint}, the well-defined form of the former (three variables and fixed coefficients) compared to the more variable nature of the latter compels us to treat them in quite different ways. While the Boros approach could in principle also be used to address constraint \eqref{const:capacity}, two significant challenges arise. First, we would need to construct the equivalent polynomial to be reduced, which in general is a highly combinatorial problem. And even so, faced with a simple case in which $c_{p}=1 \,\forall p$, where the polynomial to be penalized is simply the sum of all products composed of $C+1$ distinct $x_{p}$ variables, the number of new slack variables added scales with $\mathcal{O}(P^C)$. This makes the method unsuitable for widespread use in generic constraints.

\section{Methods}
\label{sec:methods}

In this section, we briefly describe the algorithms and solvers we use, as well as the measure of solution quality we chose for this study. Sections \ref{sec:annealing} and \ref{sec:qaoa} outline the quantum annealing process and the QAOA algorithm, respectively. In Section \ref{sec:bqm} we describe the characteristics of D-Wave's hybrid solver, and finally, in Section \ref{sec:approx_ratio} we describe the criteria we use to evaluate solution quality.

\subsection{Quantum Annealing}
\label{sec:annealing}

Quantum annealing processors are especially suitable for combinatorial optimization problems. They are thought to exploit quantum tunneling, entanglement, and superposition to escape local minima and explore a large solution space \cite{RevModPhys.90.015002}. In QA, a quantum system is made to evolve under a Hamiltonian that interpolates between a simple Hamiltonian called mixer Hamiltonian, $H_M$, of which the ground state is known, and an Ising Hamiltonian that encodes the desired problem, called cost Hamiltonian $H_C$. A usual choice for the mixer Hamiltonian is $H_M  = \sum_i X_i$, where $X_i$ is the $x$ Pauli operator on the $i$-th qubit, and $\ket{+\ldots+}$ is its ground state. On the other hand, the Ising Hamiltonian has the following form:
\begin{equation}
    H_C = \sum_i h_i\ Z_i + \sum_{i\neq j}J_{i,j} \ Z_i Z_j \ ,
\label{eq:generalIsingform}
\end{equation}
where $Z_i$ is the $z$ Pauli operator on the $i$-th qubit, $h_i$ is its on-site energy and $J_{i,j}$ the coupling between qubits $i,j$. Therefore, we have the following total Hamiltonian:
\begin{equation}
H(f(t)) = \left(1-f(t)\right) H_{M} + f(t)H_{C},
\label{eq:Adiab_alg}
\end{equation}
where $f(t) \in [0,1]$ is the interpolation function, usually defined as a polynomial function of time $t$. By means of the adiabatic theorem \cite{born1928beweis}, if we start at the ground state of $H_M$ and evolve the system slowly over time to $H_C$, it will remain at the ground state and thus provide a solution to the minimization problem.

In QA, the choice of interpolation function aims to correctly retrieve the ground state of $H_C$ in the end, but not necessarily traversing the ground state of $H(f(t))$ in intermediate steps, which allows for faster evolution.

We can minimize the corresponding function on classical binary variables $x_i$ (obtained by transforming $Z_i\to1-2x_i$). This is by analogy called energy, and can be written as:
\begin{equation}
    E(\boldsymbol{x}) = \boldsymbol x^T Q \ \boldsymbol x \ ,
\label{eq:quboQmatrix}
\end{equation}
where $Q$ is a matrix that can be assumed symmetric without loss of generality. A QUBO consists in the minimization of such functions. Current commercial quantum annealers, such as the ones provided by D-Wave, natively solve this kind of problems. The limitation to polynomials of order two comes from the fact that the Hamiltonian in \eqref{eq:generalIsingform} only couples qubits two by two, and this is intrinsic to the hardware.

Additionally, the hardware is not able to directly couple every pair of qubits, so an important feature of the quantum processing unit (QPU) is its topology, i.e. the configuration of qubit couplings. Due to this connectivity limitation, most problems require complex embedding algorithms to represent them on the QPU. Usually, several physical qubits are required to encode a logical one, creating chains that have to stay coherent, which is a challenge for current devices. For this study, two different QPUs have been used: \texttt{Advantage\_system6.4} (\texttt{Advantage}) and the recently released \texttt{Advantage2\_prototype2.2} (\texttt{Advantage2}). The topologies of these devices are Pegasus and Zephyr, respectively, the latter having higher connectivity \cite{DW-Adv64-Manual,DW-Adv2p22-Manual}.

\subsection{Quantum Approximate Optimization Algorithm}
\label{sec:qaoa}

The quantum approximate optimization algorithm \cite{farhi2014quantum} is a hybrid algorithm that combines gate-based quantum computing with classical variational-parameter optimization in order to solve combinatorial optimization problems.

This algorithm can be understood as a discretized approximation of adiabatic evolution, provided the right choice of parameters. It does so by constructing a variational ansatz that alternates $\ell$ times the unitary operators encoding the cost Hamiltonian $H_C$ and the mixer $H_M$ (same as in QA), which results in:
\begin{equation}
   U_\ell(\boldsymbol \gamma, \boldsymbol \beta) = \prod_{j=1}^{\ell} e^{-i\beta_j H_M}e^{-i\gamma_j H_C},
\label{QAOA_ansatz}
\end{equation}
where the $2\ell$ variational parameters are $\boldsymbol \gamma = (\gamma _{1}, \gamma _{2}, \ldots, \gamma _{\ell}) $, and $\boldsymbol \beta = (\beta _{1}, \beta _{2} , \ldots, \beta _{\ell})$, with $\gamma _{j} \in [0,2\pi)$ and $\beta _{j} \in [0,\pi)$.

This anstaz is applied to an initial state $\ket{\psi_0}$, which is the ground state of the mixer $H_M$ and is generally easy to obtain. Thus we have the state:
\begin{equation}
\ket{\Psi_\ell(\boldsymbol \gamma,\boldsymbol \beta)} = U_\ell(\boldsymbol \gamma,\boldsymbol \beta)\ket{\psi_0}.
\label{eq:QAOA_state}
\end{equation}

We denote $E_\ell$ as the expected value of the Hamiltonian $H_C$ when acting on this quantum state. This expectation is given by:
\begin{equation}
E_\ell(\boldsymbol\gamma, \boldsymbol\beta) = \bra{\Psi_\ell(\boldsymbol\gamma, \boldsymbol\beta)}H_C\ket{\Psi_\ell(\boldsymbol\gamma, \boldsymbol\beta)}.
\label{eq:EV_QAOA}
\end{equation}

This quantity is minimized via a classical optimization routine. That is, we aim to find the optimal parameters:
\begin{equation}
(\boldsymbol\gamma^*, \boldsymbol\beta^*) = \argmin_{(\boldsymbol\gamma, \boldsymbol\beta)} E_\ell (\boldsymbol\gamma, \boldsymbol\beta).
\end{equation}
The solution is then obtained by measuring the state $\ket{\Psi_\ell(\boldsymbol\gamma^*,\boldsymbol\beta^*)}$ in the computational basis. In principle, this should allow to obtain an energy equal to or close to the ground state energy of $H_C$.

The positive integer $\ell$ controls how discrete the evolution is. As $\ell\to\infty$, the closer to adiabatic the evolution can be, and, in general, the better the results. However, as $\ell$ increases, so does the number of parameters and the circuit depth, making optimization more difficult and time consuming. In order to compensate for this problem, we use the parameter-fixing strategy introduced in \cite{lee2021fixedparams}. Starting from a $\ell=1$ version of QAOA with random initialization (which works for shallow circuits), the $2\ell$ parameters optimized for a $\ell$-layer QAOA are used as initialization values of a $(\ell+1)$-layer QAOA along with $\gamma_{\ell+1} = \beta_{\ell+1} = 0$.

\subsection{Leap BQM Hybrid Solver}
\label{sec:bqm}

The Leap Binary Quadratic Model Hybrid (\texttt{LeapBQMHybrid}) is a component of the Hybrid Solver Service (HSS, \cite{HSS}) offered by D-Wave, which is a collection of algorithms created by this company. In order to solve optimization problems that quantum processors are unable to directly handle, these hybrid approaches imbricate classical and quantum processing \cite{HybridDwave}. As of this writing, the solvers included in HSS allow the addressing of Binary Quadratic Models (BQM), Discrete Quadratic Models (DQM) and Constrained Quadratic Models (CQM). The reason for using \texttt{LeapBQMHybrid} over the other available solvers is that the problem dealt with in this paper is defined entirely in binary variables. 

Going deeper into the algorithm, the \texttt{LeapBQMHybrid} workflow is split into two distinct phases. To be more exact, the BQM formulation is first introduced as input into a classical front end. The solver then runs a predetermined number of parallel processing threads, each of which consists of a quantum module (QM) and a heuristic module (HM). On the one hand, HM is devoted to tackling the problem by means of traditional state-of-the-art heuristic techniques. On the other hand, QM helps HM by executing a variety of quantum queries with the objective of guiding the search towards promising regions of the search space. Eventually, QM can also improve the existing solutions through these queries. Finally, \texttt{LeapBQMHybrid} provides the user with the optimal solution found among the group of generated threads. 

It is interesting to mention that \texttt{LeapBQMHybrid} resorts to the latest D-Wave quantum computer. At the time this research was conducted, this device was the \texttt{Advantage\_system6.4}, which is made up of 5616 qubits organized in a Pegasus topology. We depict in Fig. \ref{fig:LeapBQM} the main workflow of this solver, which is based in the work published in \cite{osaba2024hybrid}. Readers interested in further details about the \texttt{LeapBQMHybrid} might peruse the D-Wave related report \cite{HybridDwave}. 

\Figure[!t]()[width=0.95\columnwidth]{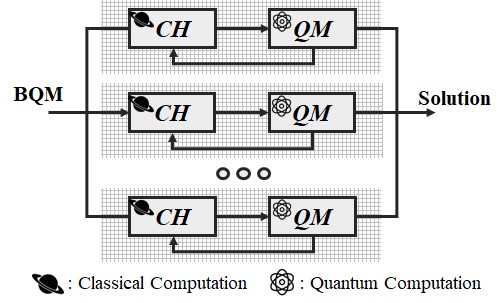}{Workflow of the \texttt{LeapBQMHybrid} solver. QM = Quantum Module. CH = Classical Heuristic Module. \label{fig:LeapBQM}}

\subsection{Solution quality assessment criteria}
\label{sec:approx_ratio}

Benchmarking the performance of quantum optimization algorithms and hardware is still an open research question without a clear answer \cite{lubinski2023optimization}. In this study, we have adopted a similar approach to prior research in the field, using the approximation ratio ($AR$) as a metric for how good a solution vector $x$ is. We define $AR$ as follows:
\begin{equation}
    AR(x) = 
    \begin{cases}
        \frac{F(x)}{F_{\max}} & \text{if } x \text{ is feasible,}\\
        0 & \text{otherwise,}
    \end{cases}
    \label{eq:appx_ratio}
\end{equation}
where a solution vector is deemed feasible if it respects the constraints in \eqref{eq:constraints}, and $F_{max}$ denotes the maximum objective function value achievable by any feasible solution.

It is important to note that from the runs of the algorithms, we get $(n+s)$-long vectors, where the first $n$ components are the actual decision variables and the remaining $s$ are slack terms introduced by the penalties. Therefore, we define the solutions $x$ to be the first $n$ components of these vectors. This definition allows us to evaluate the quality of solution vectors in the context of the optimization problem at hand, hence independently of the specifics of the QUBO encoding utilized, providing a standardized measure to assess their performance.

\section{Experimentation and Results}
\label{sec:expres}
 
This section is devoted to the description of the main elements of the experiments conducted and the results obtained. In Section \ref{sec:data} we introduce the characteristics of the dataset utilized in our study and how we generated it. Then, in Section \ref{sec:setting} we detail our design choices for the experiments performed. Finally, in Section \ref{sec:results} we discuss and analyze the results. Fig. \ref{fig:pipeline} illustrates the implemented experimental pipeline.

\Figure[!t]()[width=1\linewidth]{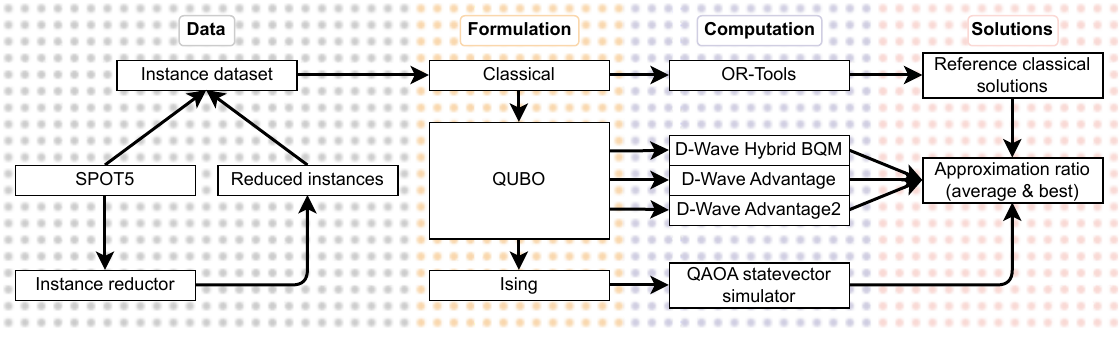}{Schematic diagram of the pipeline implemented to conduct the experiments. First, we generate the dataset by combining original SPOT5 instances with reduced ones. Then, we formulate the classical model, translate it to our proposed efficient QUBO encoding and obtain the equivalent Ising model. We get the reference solutions with OR-Tools and run the quantum/hybrid subroutines a total of 5 times per model and instance. Finally, we compute the approximation ratio for the best found solutions as well as the average from the 2000 samples of each run executed. \label{fig:pipeline}}

\subsection{Data and Instance Reductor}
\label{sec:data}

We have developed a Python script for the automatic production of SMPP instances as part of the current study. The implemented mechanism, coined \texttt{SPOTReductor}, works as follows: first, an existing instance of the problem (in SPOT5 format) is introduced to \texttt{SPOTReductor} together with the number of requests desired for the new instance. After that, \texttt{SPOTReductor} reduces the input problem by randomly selecting constraints and keeping the requests involved in them until the desired number of requests is reached. Then, the maximum capacity of the satellite is established as half of the capacity that would be used if all requests were taken. Finally, to generate the instances without capacity constraints, the capacity requirements for each request are simply set to zero and the maximum capacity is removed. This way, we can study more in-depth the influence of the capacities in the problem.

In all, the testing planned for this study has 31 distinct instances. Of them, 6 have been obtained from the SPOT5 benchmark dataset and an additional 25 instances have been created through \texttt{SPOTReductor}. With these instances, we cover a range of jobs from 3 to 209, however, it is worth noting that it is not trivial to order the instances by difficulty, as their intrinsic variety in terms of request type, amount of ternary constraints, capacities, etc. play a vital role in this regard. Here we chose to order them by the amount of requests for readability. Additionally, to improve this study's reproducibility, all generated cases along with the obtained results are openly available in \cite{dataset}.

\begin{table}[!t]
\caption{Main characteristics of the used instances, ordered by increasing number of requests. For each instance, we depict the number of total and stereo requests, the amount of total and ternary constraints, and the number of independent and interaction terms in our formulation. The naming convention is as follows: if an instance is from the original dataset \cite{bensana1999earth}, we keep its original name, which consists of a series of digits (e.g. \texttt{8}, \texttt{404}, \texttt{1502}). If the instance was generated by our reductor, we use the following pattern: \texttt{gDDD(c)} (e.g. \texttt{g003}, \texttt{g003c}), where the character \texttt{g} is used to flag that the instance was generated, then three digits represent the amount of requests in the instance, and finally, a character \texttt{c} is present for instances that have a capacity constraint.}
\label{tab:instances}
 \centering
 \resizebox{1\columnwidth}{!}{
     \begin{tabular}{lcccc | cc}
      \toprule
      ID & Requests & Stereo & Constraints & Ternary & Independent & Interaction \\
      \midrule
      \texttt{g003}  & 3   & 1  & 1   & 1   & 11  & 13    \\
      \texttt{g003c} & 3   & 1  & 1   & 1   & 14  & 81    \\
      \texttt{g004}  & 4   & 1  & 1   & 1   & 11  & 13    \\
      \texttt{g004c} & 4   & 1  & 1   & 1   & 14  & 81    \\
      \texttt{g005}  & 5   & 0  & 2   & 0   & 13  & 16    \\
      \texttt{g005c} & 5   & 0  & 2   & 0   & 15  & 27    \\
      \texttt{g006}  & 6   & 0  & 5   & 2   & 20  & 31    \\
      \texttt{g006c} & 6   & 0  & 5   & 2   & 23  & 216   \\
      \texttt{g007}  & 7   & 0  & 8   & 1   & 20  & 34    \\
      \texttt{g007c} & 7   & 0  & 8   & 1   & 23  & 234   \\
      \texttt{8}     & 8   & 4  & 7   & 0   & 16  & 29    \\
      \texttt{g009}  & 9   & 0  & 3   & 1   & 28  & 37    \\
      \texttt{g009c} & 9   & 0  & 3   & 1   & 30  & 152   \\
      \texttt{g010}  & 10  & 1  & 5   & 2   & 26  & 34    \\
      \texttt{g010c} & 10  & 1  & 5   & 2   & 29  & 155   \\
      \texttt{g015}  & 15  & 0  & 23  & 7   & 42  & 81    \\
      \texttt{g015c} & 15  & 0  & 23  & 7   & 46  & 762   \\
      \texttt{g025}  & 25  & 1  & 31  & 0   & 55  & 94    \\ 
      \texttt{g025c} & 25  & 1  & 31  & 0   & 59  & 569   \\
      \texttt{g035}  & 35  & 4  & 63  & 11  & 89  & 188   \\
      \texttt{g035c} & 35  & 4  & 63  & 11  & 94  & 1126  \\
      \texttt{g045}  & 45  & 5  & 84  & 12  & 124 & 317   \\
      \texttt{g045c} & 45  & 5  & 84  & 12  & 128 & 1412  \\
      \texttt{g060c} & 60  & 9  & 146 & 44  & 178 & 3481  \\
      \texttt{54}    & 67  & 35 & 204 & 23  & 140 & 544   \\ 
      \texttt{29}    & 82  & 29 & 380 & 0   & 120 & 667   \\
      \texttt{g085c} & 85  & 8  & 267 & 29  & 242 & 12287 \\
      \texttt{404}   & 100 & 63 & 610 & 18  & 176 & 1078  \\
      \texttt{g120c} & 120 & 10 & 529 & 155 & 379 & 16151 \\
      \texttt{503}   & 143 & 78 & 492 & 86  & 310 & 1118  \\
      \texttt{1502}  & 209 & 44 & 203 & 29  & 436 & 26304 \\
      \bottomrule        
     \end{tabular}
}
\end{table}

\subsection{Experimental Setting}
\label{sec:setting}

To conduct the experiments, we have used our QUBO/Ising formulation as described in Section \ref{sec:qubo} on the 31 instances detailed in Table \ref{tab:instances}. We obtained reference solutions (and hence $F_{\max}$) from exactly solving the classical formulation with \texttt{Google OR-Tools} \cite{ortools}. To build the QUBO matrices, we chose to set the value of all penalty coefficients in each instance to $M = \sum_i w_i + 1$, which is a standard choice used in the literature if no prior information about the instances is known \cite{glover2018tutorial}. To account for the probabilistic nature of the methods, we run each combination of instance and solver 5 times. Finally, we report the expected value of the $AR$ of the solutions and the $AR$ of the best sampled solution. To make QAOA comparable to QA in terms of the best solution, we sample from the QAOA statevector as many times as shots run for QA.

For the experiments with QA, we tested three different D-Wave solvers: \texttt{Advantage}, \texttt{Advantage2}, and \texttt{LeapBQMHybrid}. For the two QPU-only based solvers, we have left all parameters at their default values except for the number of reads, which we have set to 2000. For the hybrid solver, only the \texttt{time\_limit} parameter is tunable, which we chose to set to the default value computed by D-Wave for each instance.

For QAOA, starting with $\ell=1$ and using the parameter-fixing strategy described in section \ref{sec:qaoa}, we run the algorithm for $\ell = 1,\ldots, 10$ for each instance. For the mixer Hamiltonian, we employ the standard $H_M  = \sum_i X_i$, and thus the initial state is set to be its ground state, $\ket{\psi_0} = \ket{+\ldots+}$. We use COBYLA as the optimizer \cite{Powell1994}, configured with a tolerance of $10^{\minus6}$ as stopping criteria. We report results for $\ell = 1$ and $\ell=10$. 

To select the initial parameters from which to start the optimization process for $\ell=1$, we randomly pick for each instance 5 pairs of $(\gamma_1,\, \beta_1)$, run QAOA with depth $\ell=1$ for each of them and keep the best performing pair among the five, ranked by the approximation ratio of the output.

The executions with QAOA have been carried out in an ideal statevector simulator, namely Pennylane's \texttt{default\_qubit} device \cite{bergholm2018pennylane}. We refrained from using real hardware in the QAOA experiments as the number of requests that would need to be made to the quantum computers would be prohibitive in terms of wait times and costs.

\subsection{Results}
\label{sec:results}

Let us now analyze the outputs from our experiments, which are summarized in Fig. \ref{fig:results}. First, we remark that our classical exact optimizer achieved the optimal solutions for all instances and therefore we are not going to focus on the comparison of the quantum methods against it, but rather use these solutions as a reference to compare the quantum (hybrid) methods among themselves.

\Figure[!t]()[width=1\linewidth]{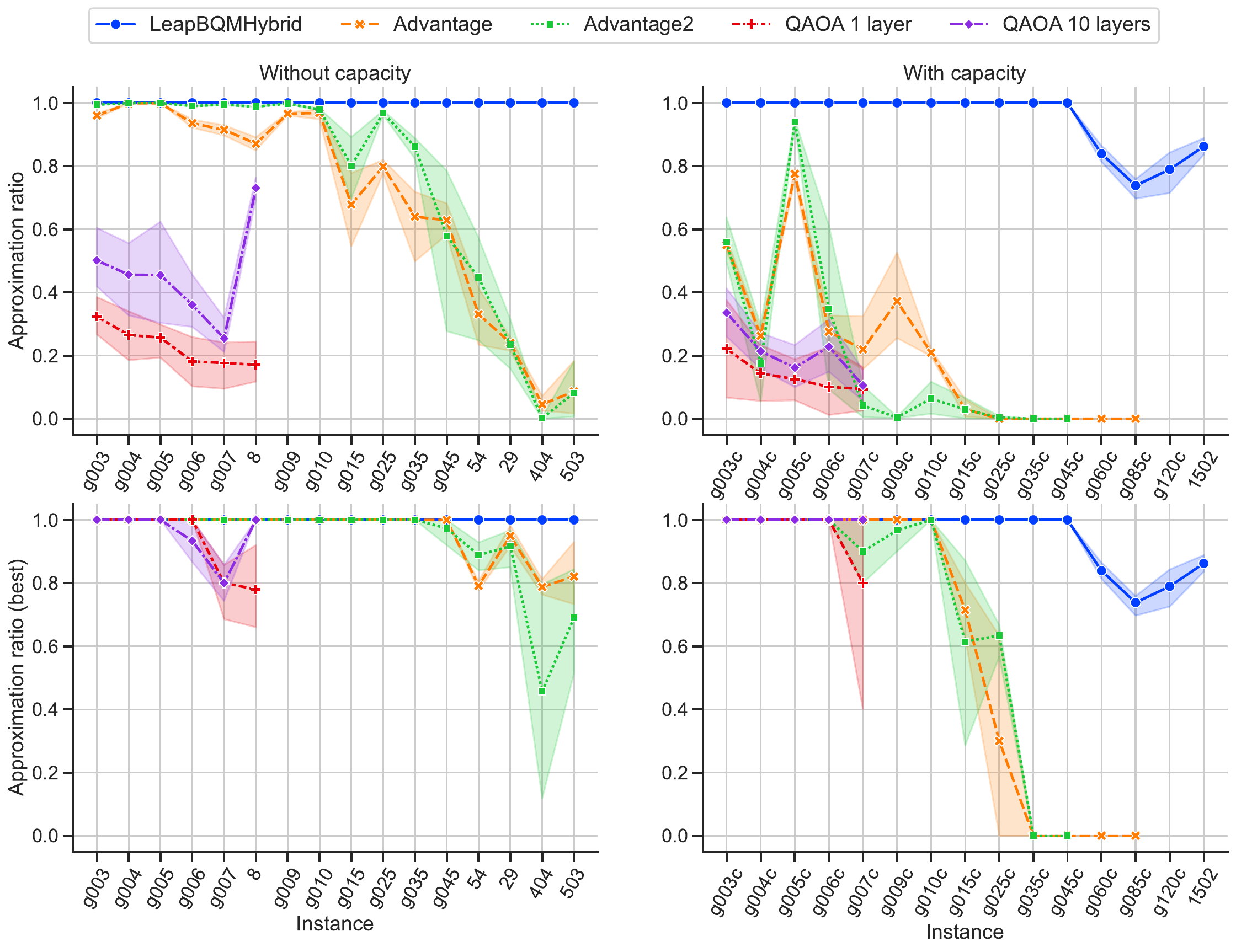}{Results for all solvers and instances considered. On the left side of the figure, we show the results for the instances without capacity constraints, while on the right side, the instances with capacities are depicted. On the top row, we represent the expected value of $AR$, while on the bottom row, the $AR$ of the best solution is shown. All plots have 95\% confidence interval error bars for the 5 runs of each experiment. Note that we did not include the classical solver's results, as their approximation ratio is always 1 without variance. \label{fig:results}}

Beginning with \texttt{LeapBQMHybrid}, we see that it performs remarkably well across the board, achieving perfect approximation ratios for all but 4 of the considered instances. However, as described in Section \ref{sec:bqm}, it is not clear how much of the processing happens on the QPU and what is its contribution. For this reason, we focus our analysis on the other methods.

In the case of the purely quantum annealing results, it is interesting to see that the \texttt{Advantage2}, while being still in development, performs better or comparably to the established \texttt{Advantage} QPU for the smaller instances. This suggests that there is an evolution in the quality of the processors and sheds a positive outlook on the problems we will be able to tackle when it is released in full scale.

As seen in the lower half of the figure, the \texttt{Advantage} solver is able to find the optimal solution for 13 of the non-capacity instances and 7 of the capacity ones, while \texttt{Advantage2} is slightly less performant, with 12 and 5, respectively. However, the general trend with \texttt{Advantage2} is that it often samples better solutions overall, as seen in the upper part of the figure, providing better average approximation ratios.

It is also noticeable that finding an embedding of a problem on the QA's QPUs does not guarantee that a reasonable solution will be found. This is evident from the larger capacity instances, where we see that from instance \texttt{g025c} onwards, even if the problem is sufficiently small to be embedded in the hardware, the approximation ratio is 0. However, this behaviour is expected: if a problem's connectivity is close to the limit of the QA hardware, the resulting embedding is very complex, necessitating very long chains of physical qubits to represent a logical variable. These long chains are more unstable and it is very challenging to keep them coherent, thus resulting in errors that ruin the solution.

Another interesting behaviour is observed in the expected $AR$ of capacity instances with QA, where we see jumps in performance even for problems with similar amounts of requests. This is due to the fact that not all requests have nonzero capacity, which is made evident in Fig. \ref{fig:graphs}, where we can see the contribution of the capacity constraint by checking the difference between the capacity and non capacity versions of the instances. We see that \texttt{g004c} is heavily connected, meaning most images need to be stored on the satellite, while \texttt{g005c} is sparse since many images can be directly relayed and thus do not add extra connections. These results indicate that QA is sensitive to the sparsity of the graph. While it might seem that QAOA is not affected by this phenomenon, we stress that we run ideal simulations, and higher connectivity still poses a problem for real gate-based quantum computers. Usually, they do not have all their qubits coupled among themselves, thus necessitating a large overhead of operations to conform to the connectivity of the problem, increasing the depth of the resulting circuits and consequently being subject to more noise induced errors. In general, the sparsity of the graph is a factor that needs to be taken into account, especially for noisy intermediate-scale quantum (NISQ) \cite{preskill2018quantum} computers and algorithms.

\Figure()[width=0.95\columnwidth]{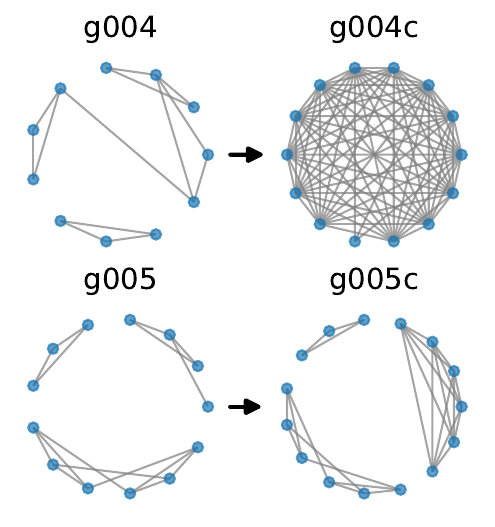}{QUBO matrices of instances \texttt{g004(c)} and \texttt{g005(c)} represented as graphs. Some capacity constraints introduce more complexity to the problem than others depending on whether or not the requests take up capacity (due to being immediately relayed or not). \label{fig:graphs}}

Turning our attention to QAOA, we see that the 10 layer version performs better than 1 layer across all instances when expected $AR$ is considered, and on all but one instance (\texttt{g006}) when comparing the best-solution $AR$. The reason a shallower QAOA might perform better can be attributed to three factors. Firstly, the stochastic nature of the sampling process. Secondly, the fact that we optimize the parameters with respect to the expected value, not the best solution. And thirdly, the parameter optimization algorithm is naturally more prone to getting trapped at local minima in higher dimensional search spaces, resulting in suboptimal parameters for circuits with more layers.

Remarkably, looking at the bottom row of Fig \ref{fig:results}, we see that QAOA performs significantly better on instances with capacities than without, which means that potentially, the graph structure influences the quality of the solutions given by the algorithm, being more suitable for complete or close to complete graphs. This finding supports the evidence from \cite{blekos2024review}, where the authors show that QAOA performs better on complete and regular graphs than on random ones. This fact might become useful in the long-term, when (if) fault tolerant computers become available, as on NISQ devices the noise mostly negates this feature.

We further propose that this phenomenon may be attributed to the fact that capacity-constrained problems naturally impose tighter constraints, resulting in a reduced number of feasible solutions. During the optimization process executed by QAOA, the algorithm tends to redistribute probability mass from non-feasible states, characterized by notably penalized energies, to feasible states. Given the scarcity of feasible states in capacity-constrained problems, there exists a higher probability, albeit fortuitously, that the redistributed probability mass aligns with the globally optimal state.

\section{Conclusions and Further Work}
\label{sec:conclusion}

In this paper, we have experimentally assessed the performance of quantum annealing and gate-based optimization algorithms on an efficient formulation of the SMPP with intricate constraints and structure. We have resorted to a realistic benchmark dataset and shown how the quality of the solutions degrades with problem size, imposing practical limits on the instances that can currently be solved effectively and providing a comprehensive overview of the challenges that arise when dealing with the SMPP in particular. This work assesses the readiness of quantum optimization methods and hardware for the SMPP, focusing on the gate-based and annealing paradigms.

We have shown that it is possible to achieve good results (best solution $AR\geq 0.9)$ for moderately sized instances of relatively complex SMPPs using quantum optimization methods. With QA, we were able to tackle instances with sizes of up to about 80 requests ($\approx 200$ variables) when no capacities are considered and up to 10 requests ($\approx 30$ variables) when capacities are included. Meanwhile, with QAOA, the sizes of the instances we are able to deal with are more modest, achieving reasonable results for instances of up to 6-8 requests ($\approx 20$ variables), both with and without capacities, although we were limited by the number of qubits we were able to simulate.

On another note, the new generation of D-Wave quantum annealers is showing promising progress, which leaves an optimistic outlook on future developments. With respect to the QAOA, even in ideal simulated environments, the performance is generally not on par with QA. However, there is evidence~\cite{Zhou2020,Wauters2020} that suggests that, compared to QA, QAOA can be more robust to vanishing spectral gaps, and it is possible that adding more layers to the circuit or finding better optimization strategies could improve the results.

Additionally, it is important to remark that the parametrization greatly influences the quality of both methods. This leads us to believe that an exhaustive tuning of the parameters of the problem (value of the penalties, variable encodings, etc.) and models (number of layers in QAOA, optimizers, annealing time in QA, etc.) might improve the results significantly, but this poses significant challenges, not only computational but also with hardware availability, especially if the tuning needs to be done on today's scarcely available quantum computers.

Lastly, future research could be focused on extending the planning to multiple satellites that can cooperate, which increases the complexity and presents a more realistic scenario of the industry's challenges. Additionally, it would also be interesting to compare our results with other variational algorithms or run the QAOA with simulated noise models or even on real hardware. Basic questions such as determining whether quantum phenomena positively affect optimization are also yet to be explored. Finally, a growing field of research is quantum-inspired optimization, where other algorithms such as tensor networks could be explored.

\bibliographystyle{ieeetr}
\bibliography{bibliography.bib}

\begin{thebibliography}{10}

\bibitem{zhang2021mission}
G.~Zhang, X.~Li, G.~Hu, Z.~Zhang, J.~An, and W.~Man, ``Mission planning issues of imaging satellites: Summary, discussion, and prospects,'' {\em International Journal of Aerospace Engineering}, vol.~2021, pp.~1--20, 2021.

\bibitem{bensana1999earth}
E.~Bensana, M.~Lemaitre, and G.~Verfaillie, ``Earth observation satellite management,'' {\em Constraints}, vol.~4, no.~3, pp.~293--299, 1999.

\bibitem{karp1975computational}
R.~M. Karp, ``On the computational complexity of combinatorial problems,'' {\em Networks}, vol.~5, no.~1, pp.~45--68, 1975.

\bibitem{vasquez2003upper}
M.~Vasquez and J.-K. Hao, ``Upper bounds for the spot 5 daily photograph scheduling problem,'' {\em Journal of Combinatorial Optimization}, vol.~7, pp.~87--103, 2003.

\bibitem{ribeiro2010strong}
G.~M. Ribeiro, M.~F. Constantino, and L.~A.~N. Lorena, ``Strong formulation for the spot 5 daily photograph scheduling problem,'' {\em Journal of combinatorial optimization}, vol.~20, pp.~385--398, 2010.

\bibitem{Wei1399860tabu}
W.-C. Lin and D.-Y. Liao, ``A tabu search algorithm for satellite imaging scheduling,'' in {\em 2004 IEEE International Conference on Systems, Man and Cybernetics (IEEE Cat. No.04CH37583)}, vol.~2, pp.~1601--1606 vol.2, 2004.

\bibitem{barkaoui2020new}
M.~Barkaoui and J.~Berger, ``A new hybrid genetic algorithm for the collection scheduling problem for a satellite constellation,'' {\em Journal of the Operational Research Society}, vol.~71, no.~9, pp.~1390--1410, 2020.

\bibitem{bianchessi2007heuristic}
N.~Bianchessi, J.-F. Cordeau, J.~Desrosiers, G.~Laporte, and V.~Raymond, ``A heuristic for the multi-satellite, multi-orbit and multi-user management of earth observation satellites,'' {\em European Journal of Operational Research}, vol.~177, no.~2, pp.~750--762, 2007.

\bibitem{jiang2022dual}
X.~Jiang, Y.~Song, and L.~Xing, ``Dual-population artificial bee colony algorithm for joint observation satellite mission planning problem,'' {\em IEEE Access}, vol.~10, pp.~28911--28921, 2022.

\bibitem{WANG20223478}
X.~Wang, J.~Wu, Z.~Shi, F.~Zhao, and Z.~Jin, ``Deep reinforcement learning-based autonomous mission planning method for high and low orbit multiple agile earth observing satellites,'' {\em Advances in Space Research}, vol.~70, no.~11, pp.~3478--3493, 2022.

\bibitem{symons2023practitioner}
B.~C. Symons, D.~Galvin, E.~Sahin, V.~Alexandrov, and S.~Mensa, ``A practitioner's guide to quantum algorithms for optimisation problems,'' {\em arXiv preprint arXiv:2305.07323}, 2023.

\bibitem{li2020quantum}
Y.~Li, M.~Tian, G.~Liu, C.~Peng, and L.~Jiao, ``Quantum optimization and quantum learning: A survey,'' {\em IEEE Access}, vol.~8, pp.~23568--23593, 2020.

\bibitem{kadowaki1998quantum}
T.~Kadowaki and H.~Nishimori, ``Quantum annealing in the transverse ising model,'' {\em Physical Review E}, vol.~58, no.~5, p.~5355, 1998.

\bibitem{farhi2014quantum}
E.~Farhi, J.~Goldstone, and S.~Gutmann, ``A quantum approximate optimization algorithm,'' 2014.

\bibitem{kaltenbaek2021quantum}
R.~Kaltenbaek, A.~Acin, L.~Bacsardi, P.~Bianco, P.~Bouyer, E.~Diamanti, C.~Marquardt, Y.~Omar, V.~Pruneri, E.~Rasel, {\em et~al.}, ``Quantum technologies in space,'' {\em Experimental Astronomy}, vol.~51, no.~3, pp.~1677--1694, 2021.

\bibitem{vallone2015experimental}
G.~Vallone, D.~Bacco, D.~Dequal, S.~Gaiarin, V.~Luceri, G.~Bianco, and P.~Villoresi, ``Experimental satellite quantum communications,'' {\em Physical Review Letters}, vol.~115, no.~4, p.~040502, 2015.

\bibitem{liao2017satellite}
S.-K. Liao, W.-Q. Cai, W.-Y. Liu, L.~Zhang, Y.~Li, J.-G. Ren, J.~Yin, Q.~Shen, Y.~Cao, Z.-P. Li, {\em et~al.}, ``Satellite-to-ground quantum key distribution,'' {\em Nature}, vol.~549, no.~7670, pp.~43--47, 2017.

\bibitem{MULLER2020105110}
F.~Müller, O.~Carraz, P.~Visser, and O.~Witasse, ``Cold atom gravimetry for planetary missions,'' {\em Planetary and Space Science}, vol.~194, p.~105110, 2020.

\bibitem{combarro2023practical}
E.~F. Combarro, S.~Gonz{\'a}lez-Castillo, and A.~Di~Meglio, {\em A Practical Guide to Quantum Machine Learning and Quantum Optimization: Hands-on Approach to Modern Quantum Algorithms}.
\newblock Packt Publishing Ltd, 2023.

\bibitem{ORBi-eceda854-1443-4d98-8492-a10520f1a2d3}
P.~Dubey and A.~Hein, ``Satellite routing with quantum annealing: Collecting space debris and on-orbit servicing,'' in {\em Proceedings of the International Astronautical Congress}, International Astronautical Federation, 06 October 2023.

\bibitem{bass2018heterogeneous}
G.~Bass, C.~Tomlin, V.~Kumar, P.~Rihaczek, and J.~Dulny, ``Heterogeneous quantum computing for satellite constellation optimization: solving the weighted k-clique problem,'' {\em Quantum Science and Technology}, vol.~3, no.~2, p.~024010, 2018.

\bibitem{stollenwerk2020image}
T.~Stollenwerk, V.~Michaud, E.~Lobe, M.~Picard, A.~Basermann, and T.~Botter, ``Image acquisition planning for earth observation satellites with a quantum annealer,'' {\em arXiv preprint arXiv:2006.09724}, 2020.

\bibitem{stollenwerk2021agile}
T.~Stollenwerk, V.~Michaud, E.~Lobe, M.~Picard, A.~Basermann, and T.~Botter, ``Agile earth observation satellite scheduling with a quantum annealer,'' {\em IEEE Transactions on Aerospace and Electronic Systems}, vol.~57, no.~5, pp.~3520--3528, 2021.

\bibitem{rainjonneau2023quantum}
S.~Rainjonneau, I.~Tokarev, S.~Iudin, S.~Rayaprolu, K.~Pinto, D.~Lemtiuzhnikova, M.~Koblan, E.~Barashov, M.~Kordzanganeh, M.~Pflitsch, and A.~Melnikov, ``Quantum algorithms applied to satellite mission planning for earth observation,'' {\em IEEE Journal of Selected Topics in Applied Earth Observations and Remote Sensing}, vol.~16, pp.~7062--7075, 2023.

\bibitem{quetschlich2023hybrid}
N.~Quetschlich, V.~Koch, L.~Burgholzer, and R.~Wille, ``A hybrid classical quantum computing approach to the satellite mission planning problem,'' in {\em 2023 IEEE International Conference on Quantum Computing and Engineering (QCE)}, vol.~1, pp.~642--647, IEEE, 2023.

\bibitem{peruzzo2014variational}
A.~Peruzzo, J.~McClean, P.~Shadbolt, M.-H. Yung, X.-Q. Zhou, P.~J. Love, A.~Aspuru-Guzik, and J.~L. O’brien, ``A variational eigenvalue solver on a photonic quantum processor,'' {\em Nature communications}, vol.~5, no.~1, p.~4213, 2014.

\bibitem{makarov2023optimization}
A.~Makarov, M.~M. Taddei, E.~Osaba, G.~Franceschetto, E.~Villar-Rodr{\'\i}guez, and I.~Oregi, ``Optimization of image acquisition for earth observation satellites via quantum computing,'' in {\em International Conference on Intelligent Data Engineering and Automated Learning}, pp.~3--14, Springer, 2023.

\bibitem{zhang2018solving}
Y.~Zhang, X.~Hu, W.~Zhu, and P.~Jin, ``Solving the observing and downloading integrated scheduling problem of earth observation satellite with a quantum genetic algorithm,'' {\em Journal of Systems Science and Information}, vol.~6, no.~5, pp.~399--420, 2018.

\bibitem{zhi2021variable}
H.~Zhi, W.~Liang, P.~Han, Y.~Guo, and C.~Li, ``Variable observation duration scheduling problem for agile earth observation satellite based on quantum genetic algorithm,'' in {\em 2021 40th Chinese Control Conference (CCC)}, pp.~1715--1720, IEEE, 2021.

\bibitem{lucas2014ising}
A.~Lucas, ``Ising formulations of many np problems,'' {\em Frontiers in physics}, vol.~2, p.~5, 2014.

\bibitem{albash2018adiabatic}
T.~Albash and D.~A. Lidar, ``Adiabatic quantum computation,'' {\em Reviews of Modern Physics}, vol.~90, no.~1, p.~015002, 2018.

\bibitem{glover2018tutorial}
F.~Glover, G.~Kochenberger, and Y.~Du, ``Quantum {{Bridge Analytics I}}: A tutorial on formulating and using {{QUBO}} models,'' {\em 4OR}, vol.~17, pp.~335--371, Dec. 2019.

\bibitem{Boros2002}
E.~Boros and P.~L. Hammer, ``Pseudo-boolean optimization,'' {\em Discrete Applied Mathematics}, vol.~123, no.~1, pp.~155--225, 2002.

\bibitem{RevModPhys.90.015002}
T.~Albash and D.~A. Lidar, ``Adiabatic quantum computation,'' {\em Rev. Mod. Phys.}, vol.~90, p.~015002, Jan 2018.

\bibitem{born1928beweis}
M.~Born and V.~Fock, ``Beweis des adiabatensatzes,'' {\em Zeitschrift f{\"u}r Physik}, vol.~51, no.~3, pp.~165--180, 1928.

\bibitem{DW-Adv64-Manual}
D-Wave Systems, Burnaby, Canada, {\em QPU-Specific Physical Properties: Advantage\_system6.4 (User Manual)}, 2024.

\bibitem{DW-Adv2p22-Manual}
D-Wave Systems, Burnaby, Canada, {\em QPU-Specific Physical Properties: Advantage2\_prototype2.2 (User Manual)}, 2024.

\bibitem{lee2021fixedparams}
X.~Lee, Y.~Saito, D.~Cai, and N.~Asai, ``Parameters fixing strategy for quantum approximate optimization algorithm,'' in {\em 2021 IEEE International Conference on Quantum Computing and Engineering (QCE)}, (Los Alamitos, CA, USA), pp.~10--16, IEEE Computer Society, oct 2021.

\bibitem{HSS}
{D-Wave Developers}, ``{D-Wave Hybrid Solver Service: An Overview},'' Tech. Rep. 14-1039A-B, D-Wave Systems Inc., 05 2020.

\bibitem{HybridDwave}
{D-Wave Developers}, ``{Hybrid Solver for Constrained Quadratic Models},'' Tech. Rep. 14-1055A-A, D-Wave Systems Inc., 10 2021.

\bibitem{osaba2024hybrid}
E.~Osaba, E.~Villar-Rodriguez, A.~Gomez-Tejedor, and I.~Oregi, ``Hybrid quantum solvers in production: how to succeed in the nisq era?,'' {\em arXiv preprint arXiv:2401.10302}, 2024.

\bibitem{lubinski2023optimization}
T.~Lubinski, C.~Coffrin, C.~McGeoch, P.~Sathe, J.~Apanavicius, and D.~E.~B. Neira, ``Optimization applications as quantum performance benchmarks,'' {\em arXiv preprint arXiv:2302.02278}, 2023.

\bibitem{dataset}
A.~Makarov, C.~P\'erez-Herrad\'on, G.~Franceschetto, M.~Taddei, and E.~Osaba, ``Benchmark dataset and results for the satellite mission planning problem.'' \url{http://dx.doi.org/10.17632/y3zx2fht3c.1}, 2024.
\newblock Online at Mendeley Data.

\bibitem{ortools}
L.~Perron and V.~Furnon, ``Or-tools.''

\bibitem{Powell1994}
M.~J.~D. Powell, {\em A Direct Search Optimization Method That Models the Objective and Constraint Functions by Linear Interpolation}, pp.~51--67.
\newblock Dordrecht: Springer Netherlands, 1994.

\bibitem{bergholm2018pennylane}
V.~Bergholm, J.~Izaac, M.~Schuld, C.~Gogolin, S.~Ahmed, V.~Ajith, M.~S. Alam, G.~Alonso-Linaje, B.~AkashNarayanan, A.~Asadi, {\em et~al.}, ``Pennylane: Automatic differentiation of hybrid quantum-classical computations,'' {\em arXiv preprint arXiv:1811.04968}, 2018.

\bibitem{preskill2018quantum}
J.~Preskill, ``Quantum computing in the nisq era and beyond,'' {\em Quantum}, vol.~2, p.~79, 2018.

\bibitem{blekos2024review}
K.~Blekos, D.~Brand, A.~Ceschini, C.-H. Chou, R.-H. Li, K.~Pandya, and A.~Summer, ``A review on quantum approximate optimization algorithm and its variants,'' {\em Physics Reports}, vol.~1068, pp.~1--66, 2024.

\bibitem{Zhou2020}
L.~Zhou, S.-T. Wang, S.~Choi, H.~Pichler, and M.~D. Lukin, ``Quantum {{Approximate Optimization Algorithm}}: {{Performance}}, {{Mechanism}}, and {{Implementation}} on {{Near-Term Devices}},'' {\em Physical Review X}, vol.~10, p.~021067, June 2020.

\bibitem{Wauters2020}
M.~M. Wauters, G.~B. Mbeng, and G.~E. Santoro, ``Polynomial scaling of the quantum approximate optimization algorithm for ground-state preparation of the fully connected p -spin ferromagnet in a transverse field,'' {\em Physical Review A}, vol.~102, p.~062404, Dec. 2020.

\end{thebibliography}

\EOD

\end{document}